\documentclass[a4paper]{article}

\usepackage[margin=1.5cm]{geometry} 

\usepackage{amsmath}
\usepackage{amsthm}
\usepackage{amssymb}

\usepackage[utf8]{inputenc}
\usepackage{hyperref}
\usepackage{siunitx}
\hypersetup{
	unicode,
	pdfauthor={Ariane Gayout},
	pdftitle={Turbulence enhances bird tail aerodynamic performance},
	pdfsubject={Turbulence enhances bird tail aerodynamic performance},
	pdfkeywords={bird,turbulence,wake,lift},
	pdfproducer={LaTeX},
	pdfcreator={pdflatex}
}
\usepackage{lineno}
\usepackage[normalem]{ulem}

\def\add#1{{\textcolor{black}{#1}}}    
\def\del#1{}

\usepackage[numbers,sort&compress,square]{natbib}

\usepackage{graphicx, color}
\graphicspath{{fig/}}

\usepackage{algorithm, algpseudocode} 
\usepackage{mathrsfs} 

\usepackage{lipsum}

\begin{document}

\title{Turbulence enhances bird tail aerodynamic performance}

\author{Ariane Gayout $^{1,2}$}

\date{
	$^1$ Energy and Sustainability Research Institute Groningen, Faculty of Science and Engineering, University of Groningen, 9747 AG Groningen, The Netherlands\\%
	$^2$Groningen Institute for Evolutionary Life Sciences, University of Groningen, 9747 AG Groningen, The Netherlands\\
}


\maketitle

\subsection*{Abstract}
\del{Turbulence is omnipresent in the atmosphere and a long-standing research interest for its complexity. For flight, this complexity is little understood.} When turbulence arises, airplanes struggle, birds thrive. This duality leads us to seek bio-inspiration for turbulence mitigation. In particular, birds often encounter their most intense turbulence at landing, due to the \add{atmospheric boundary-layer} and respond to it by dynamically using their tail over a wide range of spreads and angles of attack. Here, using a bio-hybrid pigeon tail, we evaluate the contribution of tail spread and turbulence to the sustained aerodynamic performance of the bird at landing, i.e. at high angle of attack and low speed. We first show that spreading scarcely impacts tail aerodynamics despite large aspect ratio variations; property conserved with increased turbulence. Yet, at constant spread angle and angle of attack, turbulence greatly enhances lift and drag production, which we link to modifications in the wake spatial \del{and temporal} structures, using proper orthogonal decomposition. The results suggest a wake instability suppression by turbulence which enhances flight control by negating innate wake perturbations. This gives novel insights into how bird tails evolved to navigate the turbulent atmosphere and may inspire engineers to develop turbulence-harnessing flight control solutions.
\\

\noindent
Keywords: {\textit{Columba livia} $|$ turbulence $|$ Stereo-PIV $|$ wake $|$ lift}
\\

\section*{Introduction}
Mitigating turbulence is a hard design challenge for aircraft and autonomous flying vehicles. For example, during the last decades, clear air turbulence increased in the Northern hemisphere \cite{Storer2017, Prosser2023, Foudad2024, Alberti2024}. At the earth's surface, the smaller scale turbulent wakes generated by rough natural canopies and urban infrastructures impede the stable flight of autonomous air vehicles \cite{Watkins2006, Measure2011, Lundstrom2012, Fisher2017}. In comparison, animals seem to fly effortless in turbulence, presumably by having evolved their flight abilities in atmospheric turbulence \cite{Liu2024}. Birds appear particularly well-adapted \cite{Lempidakis2022, Laurent2021, Lempidakis2024}. Recent findings \cite{Carruthers2010, KleinHeerenbrink2022} motivated the design of bird-inspired robots \cite{Chang2020, Ajanic2020, Ryu2020, Savastano2022, Zhang2023, Jeger2024, Sedky2024, Wuest2024, Phan2024, Murayama2023}, culminating in the development of PigeonBot II, a fully autonomous robot with a biohybrid morphing wing and tail under reflex control capable of stabilizing flight in turbulence, without a vertical tail or discrete flaps, like a bird \cite{Chang2024}. Although PigeonBot II demonstrated bird tails remain effective in highly turbulent flow, how bird tails accomplish this aerodynamically has not been determined. The aerodynamics of bird wings \cite{Johansson2024, Harvey2024, Withers1981, Lentink2007, Crandell2011, Altshuler2004, Chin2016} and tails has been characterized primarily in \add{non-turbulent inflow} conditions \cite{Thomas1996, Usherwood2005, Sachs2007, Usherwood2020, Harvey2022, Chin2016, Cheney2021}, \add{which we refer to in the following as \textit{laminar} (smooth) inflow conditions \cite{Barlow1999}\footnote{Note that \textit{laminar} refers here to the state of the incoming flow, as commonly used in aerodynamic studies \cite{Obligado2013,Talavera2017,Engels2019} and not to be confused with the laminar flow state at low Reynolds number prior to the transition to turbulence \cite{Avila2010}.}}. How birds use their tail in flight is readily observed, they harness an exceptionally wide range of spread angles and angles of attack \cite{Tobalske1996}. A canonical example of bird tail manipulation is the tucked low angle-of-attack posture of a pigeon tail during cruising (Fig. 1.A left) versus spread high angle-of-attack posture during slow flight, in particular takeoff and landing (Fig. 1.A right). 
The aerodynamic performance of bird and protobird wings and tails in \add{\textit{laminar} inflow conditions} has been studied across angles of attack $\alpha$ ranging from \SI{0}{\degree} to \SI{90}{\degree} \cite{Evangelista2014}. Most tail studies focused on lower angles \cite{Evans2002a, Evans2003, Thomas1993, Thomas1996, Maybury2001, Usherwood2005, Sachs2007, Usherwood2020, Harvey2022, Chin2016}, because bird tails operate at small angles during cruising flight \cite{Evans2002a}. The remarkable diversity in tail shape and length is explained by how tail morphology evolved under both sexual selection and, to a lesser degree, energetic pressures \cite{Norberg1995, Thomas1993, Thomas1996, Fitzpatrick1999}. Tail aerodynamic force is characterized by the lift and drag coefficients and the associated pitch, yaw and roll moment coefficients determining flight stability and control \cite{Thomas1993, Sachs2007, Chang2024, Nelson1998, Phillips2004}. Because previous tail aerodynamic force and airflow measurements were conducted primarily in \add{\textit{laminar} inflow conditions}, it is unclear how turbulence affects tail efficacy.

\begin{figure}[t]
\centering
\includegraphics[width=8.4cm]{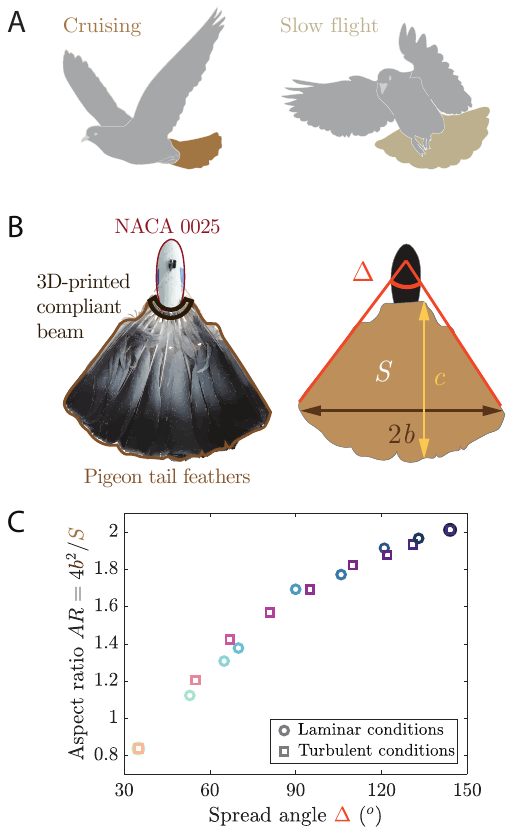}
\caption{
(A) A pigeon in cruising flight (left) with folded tail ($\alpha\simeq \SI{10}{\degree}$) and a pigeon during slow flight (right) with a wide spread tail typical for takeoff and landing ($\alpha\simeq \SI{150}{\degree}$). (B) The bio-hybrid pigeon tail with real tail feathers encased in a 3D-printed compliant strip that spreads the tail feathers (rectrices) like a fan. A NACA 0025 cylinder replaces the tail-end of the bird's body. (C) Measured biohybrid tail aspect ratio $AR$ as a function of tail spread angle $\Delta$ ($S$, surface area of the tail; $2b$, tail span; $c$, tail chord). Recordings for 1 set of racing pigeon feathers, tested under \add{\textit{laminar}} versus turbulent conditions (see methods).}\label{fig:pigeontail}
\end{figure}

To assess how turbulence affects tail efficacy, we developed a biohybrid morphing pigeon tail that we aerodynamically tested in a wind tunnel for representative spread angles of \SI{36}{\degree} to \SI{144}{\degree} and angles of attack of (\SI{0}{\degree} to \SI{90}{\degree}. \add{The choice of a pigeon tail was motivated by its squared shape, and the availability of feathers from cadavers. In particular, square-shaped tails are a common non-ornamental baseline to evolutionary studies, as it is subject to sexual selection \cite{Thomas1996,Fitzpatrick1999,Evans2004}}. The robotically controlled tail (Fig. 1.B) is a simplified version of PigeonBot II that can fly like a bird in turbulence \cite{Chang2024}. We informed the design based on the anatomy of a racing pigeon (\textit{Columba livia}) cadaver and its anatomical tail spread range that is close to natural (Fig. \ref{fig:pigeontail}.C); $\Delta = 5 \pm \SI{3}{\degree}$ (folded) to $\Delta = 148 \pm \SI{5}{\degree}$ (spread). In the new biohybrid flexure design (Fig. \ref{fig:pigeontail}.B), racing pigeon feathers are embedded in a 3D-printed flexible compliant strip that curves along a semicircle when pulled on its sides by a servo motor and spreads from \SI{36}{\degree} to \SI{144}{\degree}. This is a wider spread range than previous bio-inspired \cite{Phan2024} and biohybrid designs \cite{Chang2024}. When the tail is spread open, the change in span $2b$ and surface area $S$ (see Fig. S1) cause the aspect ratio $AR=4b^2/S$ to increase from 0.8 to 2 (Fig. 1.C). According to studies of similarly shaped rigid delta and rectangular wings in \add{\textit{laminar} inflow conditions} \cite{Winter1936, Eiffel1910}, this range of spread angles and aspect ratios can potentially change high angle of attack stall behavior.

To test the tail's aerodynamic performance, it was placed in a \textit{laminar} flow wind tunnel with a turbulence intensity $I$, below $0.5\%$ (see Methods) to measure its aerodynamic performance. To compare performance in \textit{laminar} ($I<0.5\%$) versus turbulent flow ($I \sim 8\%$) we placed a passive turbulence grid in front of the tail \cite{Mazellier2010, Melina2016, Zheng2021}. \add{Whereas this passive grid does not reproduce atmospheric turbulence due to the lack of large-scale structures, it allows us to probe the effect of a turbulence with an integral scale tuned to the tail length, which is hypothesized to have the greatest effect on aerodynamic performance of animals, independently of the turbulence rate \cite{Engels2019}}. Using robotic control, we map the effect of tail angle of attack and spread on the aerodynamic forces and wake. For each permutation, we measured the 3D aerodynamic force and moment vector using a six-axis load cell. Simultaneously, the wake was recorded using three-component (3C) velocity field measurements in the transverse wake (downstream of the tail) using ensemble stereo Particle Image Velocimetry (PIV); see Methods for details. This approach enabled us to characterize the effect of \add{\textit{laminar}} versus turbulent inflow on the aerodynamic efficacy of a pigeon tail across its operational range.



\section*{Results and discussion}


\begin{figure}[t]
\centering
\includegraphics[width=11.1cm]{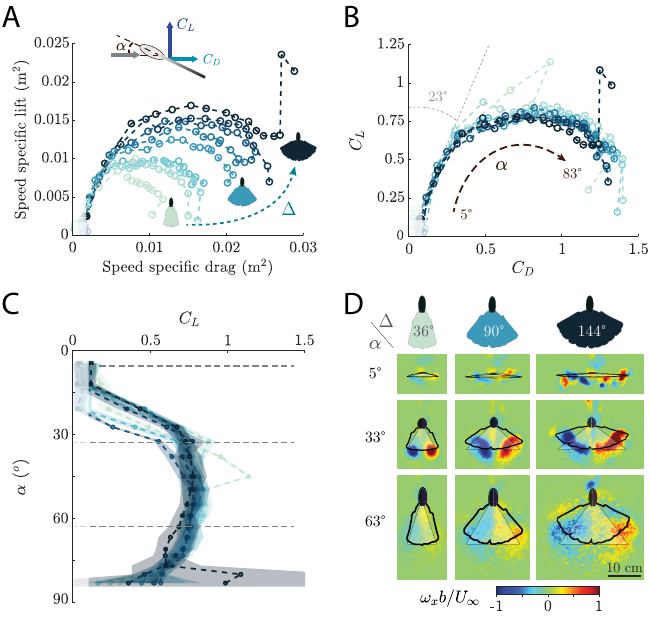}
\caption{Pigeon tail spread has little effect on aerodynamic force coefficients, which makes tail force proportional to area. (A) Polar curves of the speed specific lift $C_L S$ as a function of the speed specific drag $C_D S$ for different spread angles show force depends on tail area (semi-transparent white areas in A, B, C indicate sensor resolution limits). (B) Polar curves for the different spread angles of the lift coefficient $C_L$ as a function of the drag coefficient $C_D$ shows the coefficients depend primarily on angle of attack.  (C) Lift $C_L$ coefficients as a function of the angle of attack $\alpha$ for different spread angles. Shaded areas represent $95\%$ CI values. (D) Streamwise vorticity $\omega_x$ fields in the transverse plane behind the tail. Rows present different angles of attack $\alpha$ and columns present different spread angles $\Delta$. The projection of the tail is represented by its contour. All measurements in this figure were obtained in \add{\textit{laminar}} conditions (tail avatar color codes spread angle). 
}\label{fig:spread}

\end{figure}


\subsection*{Lift and drag are proportional to tail area}
Our experiments show that tail spread directly modulates the aerodynamic force generated by a pigeon tail through its surface area, which simplifies force control. Aerodynamic forces are proportional to force coefficient $\times$ tail area $\times$ flow speed squared. Tail lift and drag forces directly depend on tail area, which is apparent in the speed-specific lift and drag ‘polars’ (lift coefficient $\times$ tail area versus drag coefficient $\times$ tail area) (Fig. \ref{fig:spread}.A). At high angle of attack ($\alpha > \SI{50}{\degree}$), the speed-specific lift and drag are more than doubled when the tail is spread from  $\Delta = \SI{36}{\degree}$ to $\Delta = \SI{144}{\degree}$. Contrasting the speed-specific polars (Fig. \ref{fig:spread}.A) with the non-dimensional lift versus drag coefficient polars (Fig. \ref{fig:spread}.B) shows how the near-linear dependence of tail force on surface area causes the polars to practically overlap across the full spread range, \add{despite a significant variation of the aspect ratio from 0.8 to 2 (Fig. \ref{fig:pigeontail}.C)}. The overlapping coefficient polars point at an approximately constant peak lift-to-drag ratio \add{$L/D \simeq 2.4$} for all tail spreads (at $\alpha \sim \SI{23}{\degree}$). \add{This independence of the lift-to-drag ratio on the aspect ratio is remarkably contrasting with delta wings \cite{Winter1936,Gursul2005}, supporting that bird tails behave only partially similarly to delta-wings \cite{Evans2003}. This also contrasts with previous work on swifts where the change in the wings' sweep angle resulted in variations of the aerodynamic coefficients \cite{Lentink2007}}. Furthermore, during spreading, tail spread angle and tail area co-vary linearly (Fig. S1), which makes the aerodynamic forces both proportional to tail area and tail spread angle. This approximately linear relation simplifies lift and drag force control based on tail spread angle. 

The effect of tail shape, via the tail aspect ratio, on aerodynamics is primarily concentrated at low spread angles, $\Delta < \SI{50}{\degree}$ (Fig. \ref{fig:spread}.B). For a closed tail, $\Delta = \SI{36}{\degree}$, the maximal lift coefficient is about $30\%$ higher than wider spread tails at $\alpha=\SI{45}{\degree}$.  For higher angles of attack an undesired sudden large drop in lift ($\sim 20\%$) and drag ($\sim 10\%$) occurs (Fig. S2). Similar high-angle stall behavior has been reported before for various wings with equivalent low aspect ratios \cite{Eiffel1910, Winter1936} and disappears with porosity \cite{Gayout2024} \add{and elasticity \cite{Gursul2005}}. Aerodynamic porosity has been reported for bird feathers \cite{Muller1998, Eder2011}. As the tail spreads, the feather overlap reduces, which helps increase porosity effects and may thus explain why the fully closed tail is the only configuration exhibiting abrupt stall behavior. \add{Similarly, as the feather overlap reduces, the tail flexibility increases. This can further enhance its aerodynamic performance \cite{Kagawa2025}, and may mitigate the effect of aspect ratio variations, as it can be observed for flapping wings \cite{Fu2018}.}

\subsection*{Coherent tail vortices form up to high incidence}
Pigeon tails generate a coherent counter-rotating vortex pair in the wake, which is associated with delaying stall and thus continuous lift production up to high angles of attack. The angle of attack range associated with near-constant significant lift production ($C_L > 0.5$) is contained between $\alpha > \SI{30}{\degree}$ and $\alpha < \SI{75}{\degree}$ (Fig. \ref{fig:spread}.C). In this range, lift depends primarily on airspeed and spread angle, simplifying tail lift force control further, which is beneficial for high angle of attack pitch control at low speed during maneuvers and landing. This stable lift force plateau is also apparent in the dependency of vortex wake structure on angle of attack (Fig \ref{fig:spread}.D and Fig. S3), a striking strong pair of counter-rotating vortices remains continuously present at these mid-range angles of attack. This stable vortex wake structure is observed both in our bird tail recordings and previous recordings for delta wings \cite{Lee1989, Evans2003}, suggesting that geometry effects are stronger than tail morphing and porosity effects \cite{Muller1998, Gayout2024}. At low angles of attack $\alpha < \SI{25}{\degree}$, the vortex wake is diffuse with vorticity patches along the trailing edge (Fig. \ref{fig:spread}.D top row). As the angle of attack increases, a stable pair of intense coherent vortices ($\Omega_{\rm max} > U_\infty/b$, Fig. S4) develops and elongates along the trailing edge when tail spread increases (Fig. \ref{fig:spread}.D middle row). At extremely high incidence $\alpha > \SI{60}{\degree}$, \del{the tail behaves more similar to a bluff body,} as vorticity is no longer aligned with the flow direction, \add{the wake of the tail shares a structure more similar to that observed behind bluff bodies \cite{Haffner2020}}. At these extreme angles, the two vortices remain coherent but now possess a \del{turbulent-like} \add{scattered} core structure with vorticity patches spread out over a larger area (Fig. \ref{fig:spread}.D bottom row and Fig. S3). The spread-out \del{turbulent-like} \add{scattered} vortex core suggests that vortex bursting may have occurred upstream of the wake recording plane \cite{Payne1986, Kumar1998, Lentink2009a}.

A simple analysis of wake momentum transfer supports that the lift coefficient depends on tail angle of attack and not on tail spread angle, further confirming the lift coefficient and vortex wake plateau between $\alpha > \SI{30}{\degree}$ and $\alpha < \SI{75}{\degree}$. As the pair of counter-rotating vortices forms, a large downwash $w_{\rm min}$ is generated along the middle of the tail (Fig. S5). When the tail spreads, the downwash remains almost constant across spread angles ($\Delta w_{\rm min} < 30\%$). The approximate proportional scaling of similarly-shaped velocity profiles, across a large tail span (spread) range, helps explain the near constant lift coefficient $C_L$ observed (Fig. \ref{fig:spread}.C and Fig. S5). At higher angles of attack, the vorticity is no longer well-aligned with the flow direction, causing the downwash to reduce and the momentum transfer to switch from vertical to horizontal, leading to lift decrease and drag increase as observed in the polars (Fig. \ref{fig:spread}.B, \ref{fig:spread}.C and S2).


\begin{figure*}[ht!]
\centering
\includegraphics[width=17.3cm]{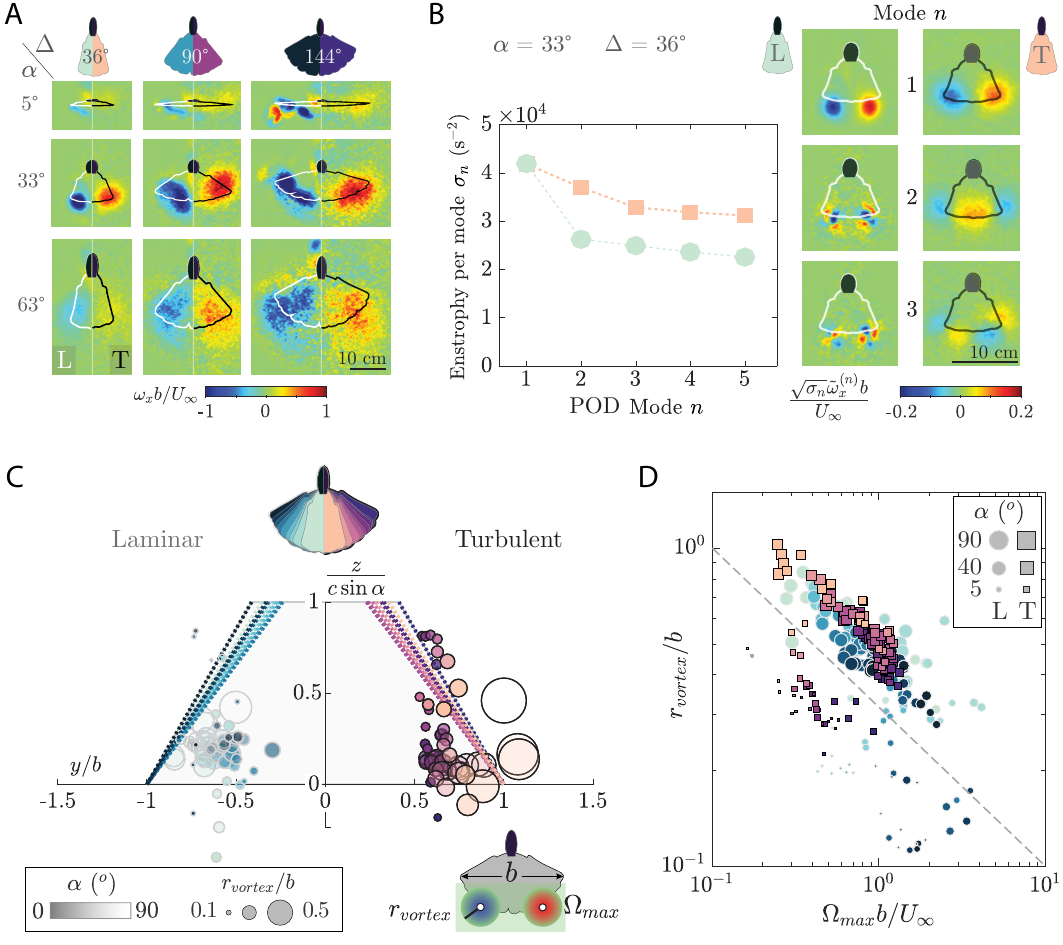}
\caption{Turbulence shifts the location of the wake while conserving circulation. (A) Streamwise vorticity $\omega_x$ fields in the transverse plane behind the tail at different angles of attack $\alpha$ (rows) and spread angles $\Delta$ (columns) in \add{\textit{laminar}} (left) and turbulent (right) conditions. Tail projection is represented by its contour. Inflow turbulence causes vortices to become rounder and secondary structures to subside. (B) Proper Orthogonal Decomposition (POD) of the streamwise vorticity $\omega_x$ fields for the configuration ($\alpha=\SI{33}{\degree}$, $\Delta=\SI{36}{\degree}$). (left) Distribution of the enstrophy across the first 5 modes for \add{\textit{laminar}} (circles) versus turbulent (squares) inflow. Enstrophy is more equally distributed across POD modes in turbulence. (right) Associated POD modes 1 to 3. Comparing \add{\textit{laminar}} (L) versus turbulent (T) conditions, the first mode shows a similar vortex structure and enstrophy. The secondary modes reveal a quadrupole vortex instability (at intermediate angles of attack) for \add{\textit{laminar}} inflow \add{conditions} that is absent for turbulent inflow \add{conditions}. (C) Location of the vortex core ($y/b$, $z/c\sin \alpha$) and radius $r_{\rm vortex}/b$ as a function of tail spread angle $\Delta$ (color) and angle of attack $\alpha$ (transparency) for \add{\textit{laminar}} (left) versus turbulent (right) inflow \add{conditions}. In turbulence, the vortices are on average bigger and located closer to the tail tip. (D) Vortex radius $r_{\rm vortex}/b$ as a function of peak vorticity $\Omega_{\rm max}b/U_\infty$ in \add{\textit{laminar}} (circles) versus turbulent (squares) conditions (marker size encodes angle of attack; 5, 40 and 90 degrees are example sizes across the full range). Both in \add{\textit{laminar}} and turbulent conditions, the trend of vortex size versus intensity follows a $-1/2$ scaling relation, associated with the conservation of circulation. The offset between low versus high angle of attack separates two regimes below and above the $-1/2$ scaling trendline.
}\label{fig:turbwake}
\end{figure*}

\subsection*{Turbulence modulates tail vortex formation}
Introducing turbulence upstream of the pigeon tail results in a more coherent wake structure with a well-developed pair of counter-rotating vortices (Fig. \ref{fig:turbwake}.A). Whereas turbulence modulates the spatial organization of the vortex wake, the coherency (Fig. \ref{fig:turbwake}.C), it does not alter the relation between vortex size and its intensity (Fig. \ref{fig:turbwake}.D). Proper orthogonal decomposition (POD) reveals the vortex enstrophy is more equally distributed across POD modes when the upstream flow is turbulent (Fig. 3.B). At low angles of attack $\alpha\leq\SI{25}{\degree}$, turbulence suppresses secondary vortices (Fig. \ref{fig:turbwake}.A top row) and diffuses streamwise vorticity into larger and less intense vorticity patches (Fig. S3). At high angles of attack $\alpha\geq \SI{50}{\degree}$ (Fig. \ref{fig:turbwake}.A bottom row), the vortex location and vorticity distribution are very similar in both \add{\textit{laminar}} and turbulent conditions (Fig. S4). At moderate angles of attack (Fig. \ref{fig:turbwake}.A middle row) however, the vortex pair is strikingly different in turbulent conditions. Turbulence causes the vortices to become larger and more circular (Fig. S4) and their profile is less skewed and more gaussian (Fig. S6).

The \del{temporal} \add{energetic} structure of the wake is strongly modulated by the turbulent inflow conditions. By analyzing the temporal evolution of the (coarse-grained) vortex wake vorticity field using POD, we obtain the temporal dependence of the spatial vortex modes as a function of the mode number $n$ and its associated enstrophy $\sigma_n$, decreasing with $n$. We found the strongest effect for $\alpha=\SI{33}{\degree}$ and $\Delta=\SI{36}{\degree}$, exhibiting the clearest differences (Fig. \ref{fig:turbwake}.B). All other configurations are available in the Supporting Information (Fig. S7-S16).  Whereas mode 1 appears structurally and quantitatively similar for \add{\textit{laminar}} and turbulent inflow conditions, the introduction of turbulence drastically modifies the structure of modes 2 and 3. The quadrupole structure (mode 2,3) inside the vortex core (mode 1) with \add{\textit{laminar}} inflow is transformed into simpler spread-out vorticity patches for turbulent inflow. For \add{\textit{laminar}} inflow conditions, the secondary vortical structures (contained in the larger vortices) across modes 2 to 10 resemble the structures reported for precessing vortices generated by delta wings \cite{Bailey2006, Dghim2021, BenMiloud2020}. The change in temporal structure of the wake thus points at a turbulent process suppressing a vortex instability of the core vortices. Enhanced vortex coherence and stability is typically associated with lift enhancement, as observed in revolving wings \cite{Lentink2009a, Nabawy2017}.

To quantify the effect of turbulence on the vortex formation, we compare vortex core location, size, and intensity. The spatial distribution of the two wake vortices in \add{\textit{laminar}} (left; blue shades) versus turbulent conditions (right; orange-purple shades) is markedly different (Fig. \ref{fig:turbwake}.C). On average, the vortex cores are closer to the trailing-edge tip of the tail in turbulence, suggesting the vortices adhere longer to the tail surface \cite{Thompson2023}. Furthermore, the size of the vortex $r_{\rm vortex}$ and its intensity $\Omega_{\rm max}$ (Fig. \ref{fig:turbwake}.D) correlate along a $-1/2$ power-law, thus $\Omega_{\rm max} r_{\rm vortex}^2$ is constant. We interpret this relation as the conservation of circulation in each of the wake vortices. The power-law fit reveals two regimes with an offset factor; low $\alpha<\SI{30}{\degree}$ versus high angles of attack $\alpha>\SI{30}{\degree}$ (Fig. \ref{fig:turbwake}.D). The separation between the two regimes corresponds to the appearance of the vortex pair in the wake (Fig. S3).  As both \add{\textit{laminar}} (blue circles) and turbulent (orange-purple squares) conditions collapse onto a common master relation that is separated by two distinguishable angles of attack regimes, this correlation also provides evidence that the introduction of free stream turbulence does not affect the vortex circulation (Fig. S4). Consequently, turbulence primarily modulates the spatial distribution of vorticity in the tail wake vortices \cite{Thompson2023}.


\begin{figure*}[ht!]
\centering
\includegraphics[width=17cm]{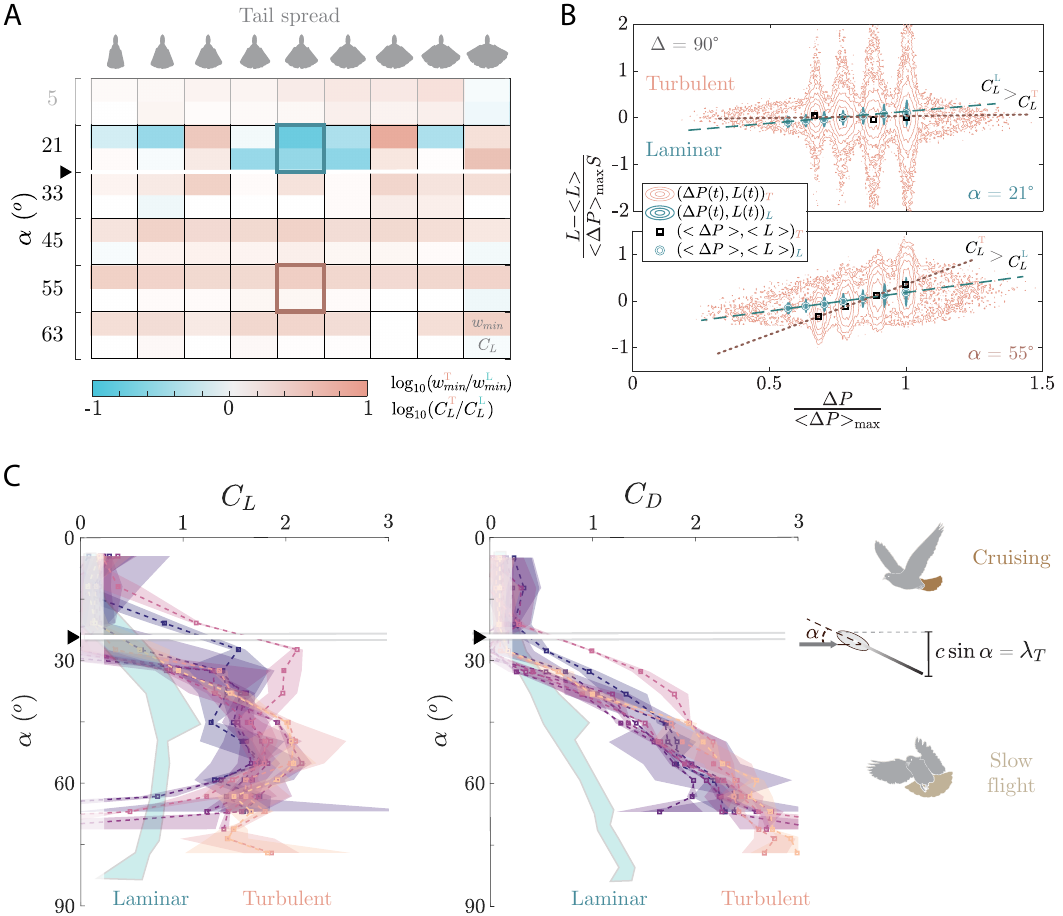}
\caption{
Tail lift and drag coefficients approximately double in turbulence. (A) Heatmap of the highest performance in average downwash velocity $w_{\rm min}$ (upper rectangle) and lift production $C_L$ (lower rectangle). The ratio of turbulent (salmon) over \add{\textit{laminar}} (teal) values are plotted for each ($\Delta,\alpha$) combination. Two regimes can be identified, one at low angles of attack where the highest performance is found in \add{\textit{laminar}} conditions and one at high angles of attack where turbulence enhances lift production. The lowest angle of attack force data is at the resolution limit of the force sensor, at $\alpha=\SI{21}{\degree}$ force data observations are supported with downwash velocity $w_{\rm min}$ measurements. (B) Distribution of the instantaneous lift ($L(t)$) and dynamic pressure ($\Delta P(t)$) measurements in \add{\textit{laminar}} (teal, darkened for more contrast) and turbulent (salmon) conditions for the spread angle $\Delta=\SI{90}{\degree}$ at $\alpha=\SI{21}{\degree}$ (top) versus $\alpha=\SI{55}{\degree}$ (bottom). Time-averaged values are represented by circles (\add{\textit{laminar}}) and squares (turbulence). Lift coefficients are correlated from the slope of the dashed (\add{\textit{laminar}}) and dotted (turbulent) lines. (C) Tail lift $C_L$ and drag $C_D$ coefficients as function of the angle of attack $\alpha$ for different spread angles in turbulence (orange to purple). The envelopes of $C_L$ and drag $C_D$ generated in \add{\textit{laminar}} conditions (teal area) are shown for reference. Shaded areas represent $95\%$ CI values. As in Fig. \ref{fig:spread}.A, tail spread angle has no clear effect on coefficient value, but turbulence doubles the lift and drag coefficients in the intermediate to high angles of attack. (Semi-transparent white areas in A, C indicate sensor resolution limit; $C_L$, $C_D$ values beyond 3 are cut-off in C). 
}\label{fig:implication}
\end{figure*}


\subsection*{Turbulence \add{can double} aerodynamic force generation} 

\add{Scale-tuned} turbulence \add{increases up to two times} the lift and drag coefficients generated by a pigeon tail at high angles of attack associated with slow flight. Force enhancement is associated with a nonlinear response of the tail to turbulent dynamic pressure fluctuations (Fig. \ref{fig:implication}). Whereas the circulation appears to be conserved when introducing freestream turbulence (Fig. \ref{fig:turbwake}.D), the intensity of the central downwafsh $w_{\rm min}$ shows clear differences between \add{\textit{laminar}} versus turbulent inflow conditions (Fig. \ref{fig:implication}.A and S5). Lift coefficient follows a similar trend (Fig. \ref{fig:implication}.A), although the low angle of attack data is close to the sensor resolution limit. For these low angle-of-attack force data observations (Fig. \ref{fig:implication}.A), additional support is provided by the roughly similar trend for downwash at ($\alpha<\SI{21}{\degree}$) and the known dependence of lift on downwash via momentum transfer \cite{Hoerner1985, Jones1990, Usherwood2020}. At moderately low angles of attack ($\alpha<\SI{30}{\degree}$), the \add{\textit{laminar}} conditions still result in a higher downwash and lift coefficient. At intermediate angles of attack ($\alpha>\SI{30}{\degree}$ and $\alpha<\SI{60}{\degree}$), freestream turbulence enhances both downwash and lift production for all spread configurations. The difference between the lift and the circulation in turbulent conditions may be interpreted as the generation of unsteady non circulatory lift \cite{Taha2020}. Comparing the distribution of instantaneous lift and dynamic pressure for turbulent inflow \add{conditions} (Fig. \ref{fig:implication}.B), shows lift fluctuations are higher at low than at high angle-of-attack. The reduction in lift fluctuation at high angle-of-attack is associated with an increased time-averaged value. This property of tails has ramifications for our understanding of bird flight control.  

The unexpected elevated performance of the pigeon tail in turbulent flow is only observed at high angles of attack beyond $\alpha_t \sim \SI{30}{\degree}$. The lift coefficient is doubled up till $\alpha \sim \SI{60}{\degree}$ after which the time-averaged lift generated by the tail stalls (Fig. \ref{fig:implication}.C). Simultaneously the drag coefficient is elevated, reaching three times the coefficient value \add{in \textit{laminar} conditions} at $\alpha \simeq \SI{45}{\degree}$ and double the value for $\alpha>\SI{60}{\degree}$ (Fig. \ref{fig:implication}.C), which enhances deceleration for landing. For low angles of attack $\alpha<\SI{30}{\degree}$, the \del{turbulent} lift and drag coefficients \add{in turbulent conditions }are of the same order as \add{in \textit{laminar} conditions} \del{coefficients}, although these low force measurements may be biased by the sensor resolution limit. The more than $100\%$ increase in lift and drag coefficient at high angles of attack (well within range of the force sensor resolution) is much larger than the increase observed in the literature, even at comparable integral length scale \cite{Ravi2012,Thompson2023}. We observe that the force elevation transition occurs at $\alpha_t$, the angle of attack for which the vertical projection of the tail $c\sin\alpha$ equals the integral length scale of the freestream turbulence $\lambda_T$ (Fig. \ref{fig:implication}.C). Identifying whether this integral length scale of turbulence causally determines the value of $\alpha_t$ would require varying the freestream turbulence with different grids. This could be a promising direction for future research into how aerodynamic performance is modulated by turbulence. If the relationship is found to be causal, $\alpha_t$ may explain bird tail fanning behavior as a function of flight speed (Fig. \ref{fig:implication}.C). Regardless, our measurements already show that turbulence enhances bird tail aerodynamic performance at high angles of attack. So whereas during cruising flight the tail is mostly horizontal with $\alpha<\SI{30}{\degree}$, as observed in pigeons, barn owls and peregrine falcons \cite{Pennycuick1968a, Durston2019, Cheney2021, Song2022}, at high angle of attack flight it is more favorable to spread the tail to deal with (un)expected turbulence. Accordingly, pigeons spread their tails at high angles of attack ($\alpha>\SI{30}{\degree}$) during slow flights \cite{Usherwood2005}, which our experiments show increases the efficacy of the tail in turbulence.

\section*{Conclusions}
To determine how turbulence affects the efficacy of bird tails at different tail angles of attack and spread angles, we measured the forces and wake generated by a robotically controlled biohybrid pigeon tail in \add{\textit{laminar}} versus turbulent \add{inflow conditions}. Our force recordings show lift and drag coefficients do not depend on \del{tail spread} \add{the aspect ratio of the tail}. \del{causing} \add{This causes the resultant forces} to be directly proportional to the tail area/spread. Wake flow measurements show how tail lift and drag forces are associated with momentum transfer in the wake and the generation of two vortical structures that are coherent in both \add{\textit{laminar}} and turbulent \del{flow} \add{conditions}. Introducing \add{freestream} turbulence modulates the formation and structure of both wake vortices, which results in an unexpected doubling of the time-averaged aerodynamic forces. Our findings suggest the effect may be explained by \del{turbulent flow} \add{ambient turbulence} suppressing a \del{laminar} vortex wake instability that \del{occurs at the low Reynolds number of birds} \add{develops in the absence of freestream flow perturbations}. Lift enhancement in turbulence has been reported before \add{for airfoils} \cite{Hoffmann1991, Ravi2012, Thompson2023, Thompson2025}, regardless the large magnitude we found for a bird tail is unexpected. The experiments suggest that bird tail efficacy \del{is} \add{can be} elevated substantially in \add{the presence of} turbulence \add{matching the integral scale to the tail}. This simplifies the mitigation of turbulent perturbations for birds and biomimetic robots harnessing biohybrid bird tails \cite{Chang2024}. \add{In particular, this matching of integral scale may be further relevant from an ecological point of view. The wake turbulence generated by the wings has a typical vortex size is about one tenth of the wingspan \cite{Usherwood2020}, which, in the case of pigeons, corresponds to the tail length.} Consequently, previous experimental measures of animal flight performance in \add{\textit{laminar}} inflow conditions cannot be simply extrapolated to explain flight stability and control under turbulent conditions. To fully explain the mechanism that causes bird tails to work more effectively in turbulence compared to their engineering counterparts, the effect of the integral turbulence length scale \add{and turbulence intensity} on the aerodynamics of lifting surfaces needs to compared at the low Reynolds number aerodynamics of birds versus high Reynolds numbers of aircraft. More immediately, a pigeon tail's linear increase in aerodynamic force as a function of tail spread angle and its elevated force generation in \add{scale-matched} turbulence informs the bio-inspired design of aerial robots with enhanced turbulence mitigation abilities.


\section*{Materials and methods}

\subsection*{Bio-hybrid pigeon tail}
For this study, we designed a hybrid robotic tail (details in \cite{Onn2023}). The tail is modeled on a pigeon (\textit{Columba livia}) with an actuated 3D printed mechanical part, corresponding to the bone and muscle tissue, and insertion of real tail feathers. The 12 tail feathers (the rectrices) were removed from a racing pigeon cadaver (N=1), purchased as animal food from a pet food supplier. The feathers were inserted into a 3D-printed compliant strip in TPU (Thermoplastic Polyurethane) that curves along a semi-circle when pulled on the sides. This compliance enabled the tail to spread in a fan-like manner by changing the traction on the strip, with a nylon string screwed on a NACA0025 cylindrical body. This cylinder was prepared with threaded holes to set the spread angle $\Delta$ at regular intervals from $\Delta=\SI{36}{\degree}$ to $\Delta=\SI{144}{\degree}$, as shown in Fig. 1.C. The robotically controlled tail (Fig. 1.B) is a simplified version of PigeonBot II \cite{Chang2024} with (i) only tail spread and pitch degrees of freedom, (ii) a single elastic connection between the feathers, and (iii) a non-covered tail feather attachment to the NACA0025 cylindrical body at the cost of some aerodynamic leakage. The NACA0025 cylindrical body also covered the pivot of the tail, to reduce aerodynamic interference of the support on the tail's aerodynamics at low angles of attack. The natural range of the tail spread angle $\Delta$ was obtained by manipulating rock pigeon (N=3) cadavers, similarly to the work on wing extensions by Stowers et al. 2017 \cite{Stowers2017} and Harvey et al. 2019 \cite{Harvey2019}.

The biohybrid tail support consisted of a steel rod (diameter \SI{8}{\milli\meter}) mounted on an ATI Nano-43 6-axis force and torque sensor (acquisition frequency: \SI{2000}{\hertz}, calibration: SI-9-0.125), together with a servomotor that controlled the angle of attack of the tail. The tail pivoted around the steel rod with ball bearings and its angle of attack was fixed by the servomotor through a steel string (diameter \SI{0.75}{\milli\meter}) encased in a carbon tube (outer diameter \SI{2}{\milli\meter}) to avoid flexion. To reduce the aerodynamic contribution of the steel rod, it was wrapped in a rigid paper airfoil to streamline it. The total aerodynamic effect of the support, including the NACA0025 cylindrical body, was measured by removing the biohybrid tail and performing force and PIV measurements at all angles of attack and flow speeds used in the experiment. Finally, angle of attack and flow speed control was automated using LabVIEW$^\circledR$ and synchronized with PIV measurements.

\subsection*{Force measurements and aerodynamic coefficients} \label{subsec:forces}
The aerodynamic forces and torques were measured using an ATI Nano-43 sensor (see Supporting Information). Only the time-averaged forces were used for the force analysis, because we do not know the cutoff frequency, determined by the natural (ringing) frequency of the force sensor setup as mounted in the wind tunnel (spectra available in Fig. S17). To reduce the effect of sensor drift, the aerodynamic forces were measured per angle of attack over six flow velocities in \add{\textit{laminar}} conditions and four to five in turbulent conditions. The difference in the number of flow velocities for which wind tunnel tests were conducted was based on the maximum force, saturation, limit of the sensor. Due to the doubling of the aerodynamic forces and unsteady force fluctuations, the sensor saturation limit was reached earlier at lower speeds (dynamic pressures) in turbulence. For each flow velocity, the acquisition was preceded with a \SI{30}{\second} wait time to ensure the flow was settled and the recordings lasted $\Delta t^L = \SI{15}{\second}$ in \add{\textit{laminar}} conditions and $\Delta t^T=\SI{1}{\minute}$ in turbulent conditions. The four-fold recording time increase in turbulence was needed to reach similar temporal convergence in the average value for the turbulent versus \add{\textit{laminar}} condition. The criterium for temporal convergence was that the standard deviation of the force mean value over averaging windows of increased duration was below $2\%$ of the standard deviation of the instantaneous force. It was obtained for $\Delta t^L > \SI{1.5}{\second}$ in \add{\textit{laminar}} conditions and $\Delta t^T>\SI{5}{\second}$ in turbulent conditions. The aerodynamic coefficients were then calculated assuming they are independent of Reynolds number over the less than one-order-of-magnitude Reynolds number range in the experiments: $\mathrm{Re}=1.2\times10^{4}$ to $\mathrm{Re}=8.9\times10^{4}$. This independence of Reynolds number is supported by similar experiments performed on delta wings \cite{Evans2002a} and protobird models \cite{Dyke2013, Evangelista2014a}. Because Reynolds number ($\propto U$) and tail deformation due to dynamic pressure ($\propto U^2$) co-vary nonlinearly with airspeed, our data cannot distinguish these effects, which motivates our approach. The aerodynamic coefficients $C_F$ were extracted by fitting the evolution of the forces $F$ (lift $L$ or drag $D$) as a function of the dynamic pressure $\Delta P$: $F(\Delta P)=a_F\Delta P + b_{\rm run}$ where $a_F=C_F S$ and $b_{\rm run}$ is a constant of the particular experimental run (Fig. S18). The confidence intervals of the coefficient values were derived from the first parameter of the linear fit (MATLAB$^\circledR$ R2022b, Curve Fitting Toolbox).

\subsection*{Flow characteristics}
The experiment was conducted in the Boundary-Layer wind tunnel at the LMFL in Lille, France. The test section dimensions are \SI{1}{\meter} height, \SI{2}{\meter} wide and \SI{20}{\meter} long. The tail was placed at the center of the test section at \SI{1.5}{\meter} from the entrance (Fig. S20.A). The nominal turbulence rate without turbulence grid was less than $0.5\%$, which we refer to as \add{\textit{laminar}} conditions throughout this study. Turbulence was generated using a grid based on a uniform $20\times 10$ grid which consists of an array of central squares integrated into a larger mesh of $\SI{36}{\centi\meter} \times \SI{36}{\centi\meter}$ (Fig. S20.E). The crossmember thickness is \SI{4}{\centi\meter} for the uniform grid and \SI{6}{\centi\meter} for the central square (Fig. S20.A,E). The integral scale of the grid $\lambda_T$ was approximately \SI{7}{\centi\meter} and the turbulence rate averaged $8 \pm 1.5 \%$  at the tail pivot location without the tail mount support present. The grid was selected based on its integral scale being of the same order of magnitude as the chord of the tail. The longitudinal position of the tail in the test section was chosen based on grid characteristics at \SI{1.45}{\meter} downstream from the grid, which we calculated to be beyond the location of peak turbulence rate \cite{Mazellier2010, Melina2016}. This location ensures the Gaussianity of the flow velocity profiles, which was verified for the instantaneous velocity fluctuations ($u,v,w$) and supports absence of large coherent structures that could otherwise bias our measurements (see Supporting Information Fig. S19).

\subsection*{Particle Image Velocimetry and Proper Orthogonal Decomposition}
To measure the three velocity components (u, v, w; 3C) in the wake of the tail, and the effect of turbulence on the wake generated by the tail, we used stereo particle image velocimetry (sPIV) in the transverse direction behind the tail at $\SI{35}{\centi\meter} \sim 2.5 c$ of the pivot (Fig. S20.B). This distance was the closest we could achieve outside the tail shadow. The sPIV system (LaVision) included a dual pulse laser (InnoLas Compact), four Imager sCMOS cameras ($f=\SI{10}{\hertz}$), and acquisition and analysis software (Davis 10.2.1, LaVision). The four cameras were set up on the same side of the wind tunnel, two facing upstream and two facing downstream (see Supporting Information Fig. S20.A). The camera setup had an angle of \SI{40.7}{\degree} with the flow direction in the tunnel. All four cameras were focused on the transverse plane, their focal plane was adjusted to align with the laser sheet using Scheimpflug mounts for the \SI{105}{\milli\meter} lenses. The laser light sheet thickness was estimated to be \SI{2.2}{\milli\meter}, with a separation of \SI{0.8}{\milli\meter}, resulting in an overlap of about $65\%$.

The stereo PIV setup records three components of velocity (3C) in the wake plane (2D) over 1000 frames recorded for \SI{100}{\second} at \SI{10}{\hertz}. The 2D3C analysis settings were preprocessed by subtracting the average background of 625 images, followed by a multi-pass vector calculation with an initial box size of 64 × 64 and a final box size of 24 × 24, with $53.4\%$ overlap (eight passes in total). The final vector field, with a vector spacing of \SI{1.2}{\milli\meter} ($\sim 8.3$ vectors/cm), was screened for erroneous vectors with a threshold at $5\%$. The final vector field spanned a maximum of \SI{41.5}{\centi\meter} in the transverse direction and \SI{40}{\centi\meter} in the vertical direction, with about \SI{2}{\centi\meter} vertical overlap between the upper and lower cameras. The vector fields were analyzed using custom written code in MATLAB$^\circledR$ R2022b. The mean vorticity fields, presented in Fig. \ref{fig:spread}.D, \ref{fig:turbwake}.A and S3, were calculated from the coarse-grained time-averaged flow fields averaged over 1000 timesteps (\SI{100}{\second}). The vorticity fields were filtered through a $15 \times 15$ px median filter (MATLAB$^\circledR$ R2022b medfilt2). To accommodate for the large variations in vorticity across the entire range of spread angles and angles of attack, the vortex radius was estimated using an adapted Q-criterion as the radius of region of $Q$ above $20\%$ of its maximal value $Q_{\rm max}$, with $Q=1/2(||\Omega_x^2||-||S_x^2||)$ being the second invariant of the velocity gradient \cite{Jeong1995}. 
To analyze the vortex spatiotemporal modes in \add{\textit{laminar}} versus turbulent conditions, a Proper Orthogonal Decomposition (POD) analysis was performed on the vorticity fields calculated from the instantaneous velocity fields, which were recorded over the total duration of each experiment and totaled 1000 images. The analysis was performed using custom code that was modified from Vincent et al. 2023 \cite{Vincent2023}. The temporal resolution window of \SI{10}{\hertz} did not allow for the identification of specific frequencies in the temporal modes of the POD. The Strouhal frequencies $f_{\rm St}$ associated with vortex shedding were estimated as being on the order of the temporal resolution of the PIV recording (10 Hz): $f_{\rm St} = St U_\infty/c$, with the Strouhal number $St \in [0.1; 2]$, $U_\infty = 2.7 - \SI{4.5}{\meter\per\second}$  and $c = \SI{15.2}{\centi\meter}$. The POD analysis of the flow field with only the tail mount present, the tail removed, confirmed that the tail mount interfered little with the tail wake (first column of Fig. S6-S10).


\subsection*{Data Availability}
Force measurements, analyzed PIV data and analysis scripts will be publicly available upon publication on the repository platform DANS. 
Due to the large size of the raw PIV data (838 GB), raw files will be shared upon request, out of logistical and ecological concerns.

\subsection*{Acknowledgements}
AG is grateful to the preliminary work and design of the experimental setup by O. Onn in his Master thesis. AG is thankful to J. C. Vassilicos and J.-P. Laval for the invitation to the Lille Turbulence Program in 2024, during which the experiments were conducted. AG also acknowledges the technical support for the experiment and particularly the PIV by P. Bragan\c{c}a. 

This publication is part of project ``On the Fly: understanding multiscale turbulence through animal flight'' (VI.Veni.232.249) funded by the Dutch Research Council (NWO) as part of the Talent Programme Veni Science domain and awarded to AG.



\bibliography{biblio_revised_260518}

\end{document}